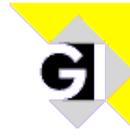

# Compliance of POLYAS with the Common Criteria Protection Profile

## A 2010 outlook on certified remote electronic voting

Niels Menke and Kai Reinhard

Micromata GmbH
Marie-Calm-Str. 1-5
34131 Kassel
Germany
n.menke@micromata.de
www.polyas.de

**Abstract:** In 2008, the German Federal Office for Information Security issued the common criteria protection profile for Online Voting Products (PP-0037). Accordingly, we evaluated the Polyas electronic voting system, which is used for legally binding elections in several international organizations (German *Gesellschaft for Informatik*, GI, among others), for compliance with the common criteria protection profile and worked toward fulfilling the given requirements. In this article we present the findings of the process of creating a compliant security target, necessary restrictions and assumptions to the system design as well as the workings of the committee, and architectural and procedural changes made necessary.

## 1 Introduction

The remote electronic voting system Polyas has been in use since 1996 in international remote electronic voting projects like the elections of the German Society for Informatics (GI), the *Deutsche Forschungsgesellschaft* (DFG), Swiss Life Group Elections, and Finnish as well as German youth elections [RJ07]. As of 2010, about a million legally binding votes have been cast using the Polyas system, supporting different methods of authentication as well as rigorous documentation while maintaining a high level of anonymity and integrity.

In 2008, the German Federal Office for Information Security and its advisory board released and certified the common criteria protection profile for remote electronic voting systems [PP08]. Since then, it has been the ambition of Polyas' developers to certify the compliance of its system and architecture with the common criteria. Toward this goal we completed a security target for the existing Polyas system based on the protection profile and adjusted the system as well as defining restrictions where necessary.

In this paper we will present the workings of Polyas and the changes made necessary to achieve compliance with the requirements of the common criteria at large and the protection profile in particular, thereby showing possible solutions to typical problems when building electronic voting systems to be evaluated against the existing common criteria protection profile.

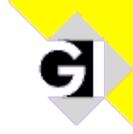

## 2 The Polyas voting process, revised

### 2.1 Overview

Polyas, among the electronic voting systems available on the market, is classified as a remote electronic voting system aka Internet voting system [VK06].
The most common variant of Polyas, which is to be discussed in this paper, uses a secret-based authentication by a common username/password process (see also [PP08] p. 16f.). While other variants of voter-authentication, namely, OpenID or Smartcard, exist and can be deployed on top of the core system, they are considered experimental at this point of time and therefore not yet to be evaluated against the common criteria protection profile.
Polyas ensures anonymity in the voting phase by means of a separation of duty among its components (see also [RJ07]). Voting with Polyas takes place by means of a Web browser (thin-client). While rich-client architecture is also available and can be used on demand of the voting committee, it is not yet subjected to common criteria evaluation.

### 2.2 Polyas general architecture—Achieving a separation of duty

The general concept of Polyas' architecture is inspired by real world ballot box voting sites (see Fig. 2.1). An electoral registry holds the authentication details and provides the point of entry for the voter who is going to cast his vote. The voter will hand his authentication credentials to the registry server, which will verify these credentials.
To ensure that the registry has not been compromised, the credentials are signed with a validation signature that resides on a third, separate validation server, and will be verified in case of authentication. Following a Two-Man-Rule, both the validation server and the electoral register will need to approve the credentials' authenticity before the voter will be issued a temporary voting token, and, with it, the opportunity to cast his vote.
To ensure that the same credentials are not used more than once for different voters (or voters unknown at the time of signing) the validator stores the signature after the first successful authentication attempt and together with the electoral register, will reject any credentials that are not eligible to cast a vote (see Fig. 2.2: Polyas Protocol).

Once the voter has received his voting token, he is passed to the ballot box server, which presents one or more virtual ballot papers for the voter to cast his vote. Once the voter has successfully cast his vote, the temporary voting token is deleted from the system, thereby destroying any link between the voter's identity and his then-cast vote.

The election process from pre-voting phase to post-voting phase is to be electronically managed and overseen by the voting committee by means of a separate system. This system will, for example, allow the committee to monitor how many votes have been cast, how many voters have been marked as having voted, oversee the system health and functionality of the other components as well as starting, stopping, and finally counting the entire vote once a configurable number of committee members has authorized each of these respective processes.

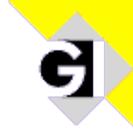

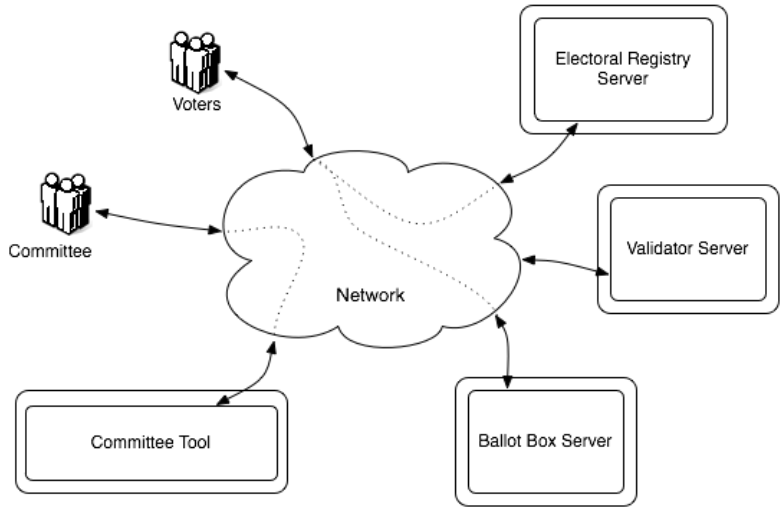

Fig. 2.1: Polyas Architecture

**2.3  Process Overview**

**Pre-Voting** There are six steps that need to be undertaken before an election can be started (See also [RJ07]):

- Installing the Polyas software on each individual server. The software should be signed to recognize software manipulation in the post-voting process.
- Generating the authentication credentials, signing them with the validators' signature, and storing them in the registry.
- Sending the authentication credentials to each respective voter. Credentials will be sent under cover and need to be revealed (a one-way-process) by the voter in order to view it.
- For each of the four Polyas components, an https, a communication, and a database key pair must be generated. The https public keys will be shared. The private communication and database keys shall be encrypted, and one pass phrase for each of the keys must be entered. These pass phrases may form an additional layer of separation of duty for the vote-starting process as they can be handed out to different members of the committee and entered separately into the committee-tool.

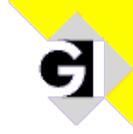

- The private communication keys of the ERS and VS are used to sign the hashed credentials of each respective voter. Let $sk_{VS}$ be the validators' communication key and $sk_{ERS}$ be the electoral registry's communication key.
  Further, let $hash$ be the SHA-256 hashing function and $sig$ be the RSA signature function.
  Then each column will contain:

  $$ID - hash(Pw) - sig_{ERS} - sig_{VS}$$

  where

  $$sig_{ERS} := sig(sk_{ERS}, hash(Pw)) \quad and \quad sig_{VS} := (sk_{VS}, sig_{ERS}).$$

  The thus a signed electoral register shall be installed on the register system. The whole electoral register is further signed with $sk_{ERS}$. This signed register should then be stored in case the need for validation arises.
- Once all components are online, the election is waiting to start. A configurable number of committee members must approve the start of the election in the committee-tool under their respective logins. Once this has happened, the system is awaiting passphrase authorization.
- For each of Polyas' components there will now be two remote access tokens (passphrases) in existence, which will have to be entered before the respective system will be operational. For the committee-tool, these shall be entered separately. When the committee tool is online and the start has been authorized, the tool will provide an interface for the committee to enter the respective passphrases of each other component.
- Once the last passphrase has been entered, the election enters the voting phase.

**Voting** The high level protocol of a voter casting a vote is described in Fig. 2.2. It is distinctive in several ways: For one, the vote is already sent to the ballot box server after the first acknowledgment. Then, the exact sent vote is sent back to the voter for verification. Thus the voter can be sure the ballot box server has interpreted his vote correctly. Votes are generally stored in an encrypted and signed manner.

Moreover, the tokens are also stored, encrypted using the public key of the involved database. Note that, according to the requirements of the protection profile, the token is explicitly not stored in the database when it is first sent to the ballot box server. It is only after the voter has confirmed his vote to be cast that the vote is finally written to the database.

Aside from the requirements of the protection profile and the signing and encryption of each individual vote, each block of thirty votes (whilst thirty is a variable) will be stored alongside with a signature of this block, factoring in the signature of the previous block, in the case that more than thirty votes have already been stored, providing a further layer of protection against any possible manipulation.

The voting token represents the authentication of the voter to the ballot box, so the ballot box cannot link the incoming or already cast vote to the credentials of the voter who issued the vote. An attacker attempting to break Polyas' anonymity would have to have unencrypted access to both of the fully separated systems (electoral register and ballot box) to establish such a link. Additionally, the vote token is encrypted via RSA, so the attacker would have to know the private key of the ballot box and/or electoral registry server in order to intercept it. Note that at no time will the token be written to the database. After the voting token is marked as invalid, its acquired memory is overwritten with pseudorandom values to ensure secure deletion.

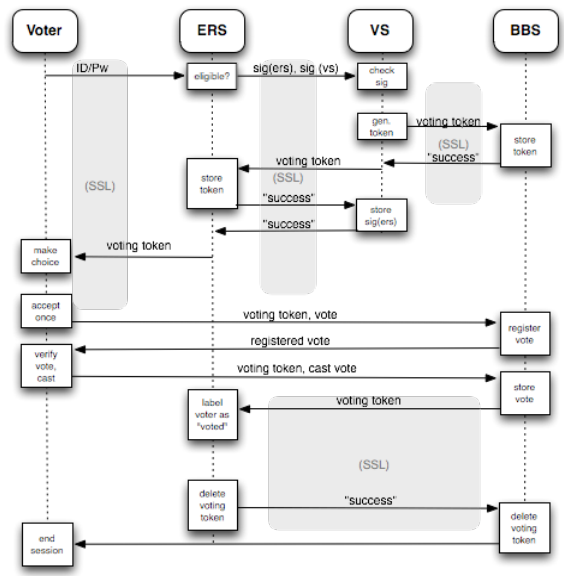

Fig. 2.2: Polyas Protocol

To provide a means of defense against so-called phishing attacks, Polyas uses a module based on Image-Maps, presenting the user with a virtual, clickable keyboard on screen. There, the user can enter his credentials and the browser will only submit X/Y-coordinates. Because these are randomized with every different login-attempt, the risk of password phishing is drastically reduced.

The protection profile requires a voter be able to cancel his voting process as well as be able to intentionally cast an invalid vote. Both requirements are fulfilled. If the voting process is intentionally cancelled or technically interrupted before the vote is committed, no vote will be stored in the database, and the voter will be able to vote again. If the process is interrupted once the vote has been finally cast, the voter will be notified of his cast on his next login-attempt.

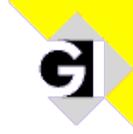

We consider the protocol to be safe against the voter trying to sell his vote. The possibility of selling would imply there being proof of his vote (and its content). Aside from the so far not solvable dilemma of remote voting, namely, that the voter can be observed throughout the entire voting process (see [KV05]), it is not possible for the voter to review his vote once cast. Therefore, it is also not possible to prove the contents of his vote to a third party after having submitted the vote and/or before submitting, since the voter might always change his choice shortly before finally casting his vote. Once the vote is cast, the voter will only be presented with the message that he has indeed voted, but for said reason no further details on his vote will be given.

**Post-Voting** To close the election, the committee has to issue the command to stop in the committee-tool. Once a sufficient number of "stop election" commands to satisfy the separation of duty has been cast, the committee-tool automatically walks through the process of stopping the election (see also [Me08]). For this purpose the validator server is first taken offline; thus disabling the possibility of further logons but not disturbing any possibly still ongoing vote processes.
After a certain amount of time to allow any remaining logged on voters to cast their votes, e.g., ten minutes, has expired, the electoral registry server is also taken offline, thus cancelling any eventually ongoing vote processes.
The ballot box server is then issued a command to count the votes and store the result along with a signature as a certificate of authenticity. The signed result can be retrieved by the committee from the ballot box server and is also displayed in the committee-tool. The committee-tool further generates a post-voting documentation including the results of the count, the log files of all involved systems, an image of each respective database, and the electoral register. All of this data will be stored in a signed archive.
As the software has been signed in the pre-voting process, it should be verified in the post-voting process that the software is still carrying the same signature to exclude the possibility of unauthorized modifications.

## 3 Achieving and maintaining compliance

### 3.1 Assessing the challenges

The Polyas architecture and process as described in 2.2 and 2.3 already fulfilled many of the objectives presented by the common criteria protection profile [PP08], as was already suggested in [RJ07]. The practicality of the implemented solutions for non-political remote electronic voting had been proven as mentioned in the introduction and in [VK06].
For one example, the protocol used by Polyas offers a natural way of achieving secrecy and anonymity when voting by fully separating the systems responsible for authenticating the voter and receiving/tallying his vote and only maintaining linkage in the form of a secure token that will be deleted at the very moment the voter has cast his vote. Simultaneously, the objective to only allow legit voters, who are unmistakably identified, had already been achieved, as was the secrecy and integrity of messaging, and the so-called overhaste protection that ensures that a voter will not cast an irreversible vote in error.
There were however, unfulfilled requirements given by [PP08] concerning the handling of the committee's tasks and its separation of duty, as well as preventing the tallying of intermediate results by members of the committee.

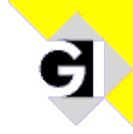

### 3.2 Assumptions and strict conformance

The common criteria protection profile for remote electronic voting does make certain conditions about the operation of the voting system that may not be circumvented for the certificate to remain valid. These conditions include, among others (for a complete list, please see [PP08]):
- The initial data in the electoral registry is that which the committee has approved. No additional data is entered by any means.
- Every registered voter has successfully received his credentials.
- The surrounding technical environment and network will function correctly for the time of the election.
- The voter will not be observed while voting (see 2.3 on vote buying).
- The committee can be trusted and will only use the functionality provided by the target of evaluation.
- The voter will verify he is connected to the correct voting system before voting.
- Data that is not under the control of the target of evaluation will be deleted once the vote has been successfully cast.

These assumptions reduce the functionality to be implemented to achieve compliance to a subset that is provided exclusively by the Polyas system as the target of evaluation.
Additionally, the protection profile demands strict conformance, which essentially means that all of the requirements have to be fulfilled by the target of evaluation itself (here: Polyas) and not by any organizational means 'on top' of the actual software.

*[Margin comment: Niels Menke 7.6.10 17:00 — Gelöscht: commands]*

### 3.3 Achieving separation of duty for the committee

One of the main challenges presented by the protection profile was the implementation of strict separation of duty for the election committee. This has been achieved in Polyas by introducing a fourth system to the original three systems in [RJ07], encapsulating the full functionality that the election committee can and may use to administer and oversee the election. This is supported by the assumption that the committee is to be trusted to not use any other knowledge or method to manipulate the election (see 3.2).
The aforementioned system, the Polyas committee-tool, integrates smoothly into the Polyas election lifecycle. It allows for the committee to safely, easily, and traceably start and stop as well as count and archive the complete election. In addition, it allows the committee to oversee the election, monitor the state of every involved system, runn self-tests, view the logging of all involved systems, and see how many votes have been cast up to the point of examination as well as how many voters have been marked as having voted. Anomalies in this case can thus easily be detected even while the election is still in an ongoing state, so the committee could decide upon measures to be undertaken in case of any discrepancies.

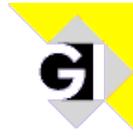

When the election is to be counted, the committee-tool provides the option of warning the committee if the number of cast votes falls below a configurable amount, thereby possibly endangering the anonymity of the cast votes.

The most prevalent feature of the committee-tool, though, is its rigorous enforcing of the separation of duty for the committee. For every election, the separation of duty count variable $S$ with $S > 1$ may be configured to a size appropriate for the specific committee.

The system will then only execute the functions of starting, stopping and/or counting the vote once $S$ different committee members have authorized this particular function with their respective credentials.

Once a committee member has given his or her authorization for a task (i.e. starting the voting phase), the committee-tool will inform him on the number of additional authorizations needed until the requested action will be carried out by the committee-tool. Every committee member may, of course, authorize each action once and only once.

### 3.4 System Safety and Self-Testing

The protection profile states that the election officers must be notified of malfunctions of the network connection or of storage of data. In such cases, the election officers should carry out a test sequence provided by the target of evaluation as demonstration of the correct operation (self-test) [PP08].

This requirement was achieved by including an already mentioned self-test routine in the committee-tool. This routine can either be carried out manually on request of an election officer, whereby it is ensured that only one self-test routine can be issued at once in case of multiple logged in election officers at the same time, or can be configured to run on a time-based schedule. In case of any detected faults at the levels of each system's hardware, storage integrity, system-time, anomalies in number of cast votes or network connection, the committee will immediately be notified of the fault and any possible consequences for the election and be asked to take appropriate counter-measures.

Any noticeable problems during the aforementioned self-test routine will be logged alongside with timestamps and therefore be included in the election archive documents.

### 3.5 Prevention of intermediate results

The protection profile requirement that no information flow between the committee and the ballot box server may result in intermediate results to be extrapolated in any way ([PP08]). Because the protection profile is formulated under the assumption that (see 3.1) the committee will only use the means provided by the target of evaluation itself, and because the committee usually will not have any direct access to the ballot box server, restricting the acting possibilities of the committee during the voting phase can solve this.

Once the vote has been started, there is no possibility offered in Polyas for the vote to be tallied unless the election is also stopped in the process. While the committee may oversee how many votes have been cast at every point in time during the voting phase, no disclosure on the content of these votes is ever given before the vote is finally stopped. Note that once stopped, in accordance to the protection profile, the election may not be resumed. Restarting a stopped election will unavoidably require the ballot box server to be cleared of any votes that had so far been cast.
Further, the stopping of the election as well as an assumed restart would have to be authorized by each of the $S$ members of the committee, hence would not go unnoticed by at least $S$ members of the committee as well as the voters who will be trying to vote during the—should such an attempt be made—inevitably resulting down-time of the voting system.

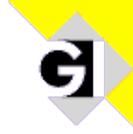

### 3.6 Audit records for the committee

The protection profile requires the committee to be able to read the audit information (successful identification and authentication of election officers, starting and stopping of the polling phase, starting of the tallying with determination of the election result, performance and results of every self-test and identified malfunctions) from the audit records of each involved system [PP08]. This information is made available in Polyas by means of the committee-tool, where each committee member can inspect the logs of each of the four Polyas component-systems in an easily readable and comprehensible format. Note that these audit files explicitly do not contain any information on the voters' logins, the identities of voters who have or have not cast their vote nor any vote content so no conflict arises with the given security objectives, particularly the secrecy of voting.

## 4 Conclusion

In this paper, we presented possible solutions to the challenges presented by the common criteria protection profile for remote electronic voting systems using the example of the Polyas system. The first look in respect to the then upcoming protection profile in 2007, [RJ07], still presented some challenges to overcome regarding the compliance of a state-of-the-art electronic voting system to the requirements of the common criteria protection profile. Additionally, there was no proof of the practicality of [PP08] so far.
The final version of the protection profile, by implying strict conformance, made organizational solutions a non-option. Instead, each requirement of the protection profile had to be directly implemented into the voting system. To achieve compliance for the Polyas system, certain minor adjustments to the protocol were necessary; as was a new tool for the committee to restrict its action options, its monitoring of the voting system's health, its view of the audit records, to enforce a separation of duty among committee members, and to prevent the tallying of intermediate results. As has been described, all of those objectives could be fulfilled while still maintaining strict conformance as well as preserving the advantages of the originally implemented protocol concerning secrecy of voting and the one-voter one-vote principle. An architectural balance between anonymity and security is still maintained in a sufficient manner for non-political remote electronic voting.

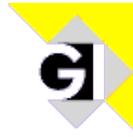

At present, we consider the described system to be compliant with the current protection profile and are looking toward qualified evaluation to achieve independently certified remote electronic voting. Therefore, we are confident that we have shown that it is possible to implement an electronic voting system for non-political voting systems that fulfill the criteria given by [PP08].

The [PP08] certification will be the first of its kind in the world of pc-based remote voting. The common criteria process will assure consistent and trusted evaluation, as well as opening up possibilities to further build upon attained knowledge and extend the acquired solutions. We look forward to additional challenges presented by the certification and publishing the first practical common criteria security target based on the protection profile.